\documentclass{article}
\usepackage{emulateapj}
\usepackage{apjfonts}
\usepackage{timesfonts}
\usepackage{psfig}
\begin{document}

\title{On the Structure of Quasi-Universal Jets for 
Gamma-Ray Bursts}
\author{Nicole M. Lloyd-Ronning$^1$, Xinyu Dai$^2$, Bing Zhang$^2$}
\affil{$^{1}$Los Alamos National Laboratory, MS B244, Los Alamos NM, 
87544; lloyd@cita.utoronto.ca \\
$^{2}$Department of Astronomy \& Astrophysics, Pennsylvania 
State University, University Park, PA 16803; 
xdai@astro.psu.edu,bzhang@astro.psu.edu\\}

\begin{abstract}
 The idea that GRBs originate from uniform jets has been used to explain 
numerous observations of breaks in the GRB afterglow lightcurves.
 We explore the possibility that GRBs instead originate from a structured
jet that may be quasi-universal, where the variation in the observed
properties of GRBs is due to the variation in the observer viewing angle.  
We test how various models reproduce the jet data of Bloom, Frail, \&
Kulkarni (2003), which show a negative correlation between the isotropic
energy output and the inferred jet opening angle (in a uniform jet
configuration).  We find, consistent with previous studies, that a
power-law structure for the jet energy as a function of angle gives a good
description. However, a Gaussian jet structure can also reproduce the data
well, particularly if the parameters of the Gaussian are allowed some
scatter. We place limits on the scatter of the parameters in both the
Gaussian and power-law models needed to reproduce the data, and discuss
how future observations will better distinguish between these models for
the GRB jet structure.  In particular, the Gaussian model predicts
 a turnover at small opening angles and in some cases a sharp cutoff at
large angles, the former of which may already have been observed. We also
discuss the predictions each model makes for the observed luminosity
function of GRBs and compare these predictions with the existing data.
 
\end{abstract}

\section{Introduction}  
  One of the outstanding problems in the field of GRBs is understanding
the extent to which these events are beamed, as well as the structure or
configuration of the jet that produces the burst and its afterglow.  The
simplest model is a uniform jet, in which it is assumed that all
parameters (density $n$, Lorentz factor $\Gamma$, magnetic and electron
equipartition factors $\epsilon_B$ and $\epsilon_e$, etc.) are constant
throughout the jet.  This jet can be described by an opening angle
$\theta_j$ (or alternatively a solid angle $ 2\pi(1-\cos\theta_j) \sim \pi
\theta_j^2$,).  Under this assumption of uniformity throughout the jet,
Frail et al.  (2001) inferred the jet angle to a sample of GRBs with
observed afterglows, from a break observed in the afterglow light curve.  
In a uniform jet model, this break occurs about the time when the GRB
ejecta has slowed down enough so that the relativistic beaming angle of
the radiation, $\sim 1/\Gamma$, becomes greater than $\theta_j$ (e.g. see
Rhoads, 1997, Frail et al., 2001).
 Using the inferred jet angle, the measured flux and redshift to each
burst, Frail et al. (2001), and then Bloom, Frail, \& Kulkarni (2003; hereafter BFK) with
a larger sample, were able to determine the emitted energy $E$ of each 
burst.  
Remarkably, they found that the GRBs in their sample exhibited very little
dispersion in $E$ (see Figure 1 of BFK); in
other words, they found that the isotropic equivalent energy of the GRB,
$E_{iso}$, and the jet opening angle, $\theta_{j}$, adhere to the
relationship $E_{iso}\theta_{j}^{2} \sim$ constant.  This intriguing
result lends some credence to the possibility that a uniform description
of a GRB jet may be valid.

   However, an alternative model was suggested by Rossi et al. (2002) and
Zhang \& Meszaros (2002; hereafter ZM).  They suggested that in fact GRBs
may have a structured jet configuration, in which parameters such as the
emitted energy can vary as a function of angle from the jet axis. All GRBs
may then have approximately the same (``quasi-universal'') jet profile but
appear different because of different orientations of the GRB jet to the
observer (or in other words, varying observer angle $\theta_{v}$).  In
this model, a break in the afterglow light curve is still observed at the
time the Lorentz factor slows to a value $\Gamma \sim 1/\theta_{v}$ (see
Rossi et al., 2002, and ZM for more discussion).  There are several
advantages of this model.  It has the appeal that it allows for uniformity
among the GRB population, and also makes definite predictions about the
distribution of observed break times in the GRB afterglow light curve
(Perna, et al. 2003), as 
well as the observed GRB luminosity function (ZM).  
The uniform jet model has to require different bursts having different
opening angles,
 and it has no predictive power for the observed distribution of these
opening angles or the
 GRB luminosity function.  In addition, realistic simulations (e.g. W.
Zhang, Woosley \& MacFadyen, 2002) naturally predict jet structure from
the collapsar scenario, so it is essential to investigate the possible jet
structures in the quasi-universal picture.  Motivated
by the Frail et al. result ($E_{iso} \theta_j^2 = $constant), the most
straightforward universal jet model would be a power-law with an index of
$k=-2$, as Rossi et al. and ZM have discussed.  However, ZM observed that
in terms of interpreting the lightcurves and jet breaks, one does not need
to abide by any power-law jet configuration. In particular, they suggested that a
Gaussian structure (and more general configurations) may describe the GRB
jet. It has also been shown that a simple power-law model 
violates some of the
 observational data.  For example, Granot and Kumar (2003), found that
this type of model
 cannot reproduce the observed afterglow light curve, while Kumar and
Granot (2003) have found that a Gaussian structure for the universal jet
model does a better job.  

  In this paper, we test how well the quasi-universal jet model can
reproduce the existing data.  In particular, we test power-law and
Gaussian models for the emitted energy, $\epsilon(\theta)$, as a function
of angle from the jet axis.  Because it is unphysical to expect that all
GRBs have exactly the same jet structure, we allow for realistic scatter
in the parameters of the models and place limits on this necessary
scatter.  The paper is organized as follows:  In \S 2, we describe the
data and any possible selection effects that may play a role in the
results.  In \S 3, we show how pure power-law and Gaussian models fit the
data of BFK.  In \S 4, we introduce scatter
into the parameters of each model and show how much dispersion is needed
in these parameters to accurately reproduce the data.  We also discuss predictions the
Gaussian model makes in the $E_{iso}-\theta_{j}$ plane, which may be tested with future
observations (and may have already been observed).  In \S 5, we discuss
what each model predicts for the observed luminosity function and how this
compares to existing data.  A summary and conclusions are presented in \S
6.

\section{Data}
  Our data is taken from BFK, which 
provides an isotropic energy estimate for 27 bursts.  Of these bursts, 16 
have a clear break in the afterglow light curve from which a jet opening 
angle (in the uniform jet model) can be inferred.  These
 data are shown in  Figure 1 (square points).  Of the 11 bursts eliminated from the 
original sample, 8 have upper (or lower) limits to $\theta_j$, based on a 
steeper (shallower) light curve observed at later (earlier) times, but no 
break directly observed.  The other three bursts were well described by a 
single power-law throughout the extent of the observations. 
As discussed in the introduction, they found in this sample a very small dispersion
for the energy emitted from a uniform jet of opening angle $\theta_{j}$.  This translates
to a negative correlation between the isotropic equivalent energy $E_{iso}$ and the jet opening
angle, such that $E_{iso}\theta_{j}^{2} \sim$ constant.  We performed
all of our analysis below on both the nominal 16 bursts as well as the sample 
where the 8 limit bursts are included (where we took the limit as the 
nominal value), which gives us some idea of how eliminating these bursts 
biases the sample and our results.  We find in fact the results do not 
change qualitatively or quantitatively in either case, so we consider our sample 
with the limits removed reasonably complete.
We present results for the sample of 16 bursts for which a definite jet opening angle has been
estimated.

   Of course, there could be additional underlying selection affects
biasing the entire sample that must be considered.  This is a concern, for
example, if there is a selection against detecting high $E_{iso}$, high
$\theta_{j}$ bursts or low $E_{iso}$, low $\theta_j$ bursts.  Such
selection effects could produce an {\em artifical} correlation in the
$E_{iso}-\theta_j$ plane.  This latter bias could be particularly
worrisome.  First, a low $\theta_j$ implies an early jet break time, and a
steepening of the light curve that may lead to a missed afterglow
detection.  More importantly, however, is that a low $E_{iso}$ may be
missed due to the flux limit of the detector.
 For a burst at some redshift $z$, the
limit on $E_{iso}$ is given by $E_{iso,lim} = F_{lim}4 \pi d^{2} (1+z)$,
where $F_{lim}$ is the limiting fluence (time integrated flux) of the
detector and $d$ is the metric distance as a function of redshift.  For
bursts at a redshift of $1$, for example, we find that $E_{iso,lim} \sim
10^{50} (F_{lim}/3\times 10^{-8}{\rm erg cm}^{-2})$ erg.  This is well below
the values of $E_{iso}$ in our sample. [We also point out that there is
some evidence that bursts have higher $E_{iso}$ at higher redshifts
(Lloyd-Ronning, et al. 2002, Amati et al., 2002, Yonetoku et al. 2003,
Graziani, et al. 2003),
 which helps reduce the severity of the flux selection.] In addition, the
expression for computing $\theta_{j}$ (see equation 1 of Frail et al.,
2001) is very weakly dependent (to the $1/8$ power) on $E_{iso}$.  This
means that the functional form of the truncation in the $E_{iso} -
\theta_{j}$ plane goes as $E_{iso} \propto \theta_{j}^{-8}$, which is far
steeper than the correlation observed.  As a result, we conclude that
selection effects are probably not producing the correlation between
$E_{iso}$ and $\theta_{j}$ in the BFK data, and that this is a reflection
of a real physical effect in GRBs.  Under the assumption that this
correlation is real, we explore configurations of the energy as a function
of angle $\epsilon(\theta)$ in the quasi-universal configuration.

 \section{Quasi-universal Jet Models} 
  In the  universal jet model, the oberver angle $\theta_{v}$ takes
the place of the jet opening angle $\theta_{j}$.  Then - as described in
ZM - $E_{iso}(\theta_{j})$ in the uniform jet model
translates to $\epsilon(\theta)$, the energy as a function of angle from
the jet axis, in the quasi-universal picture.  In principle, we can then
fit $E_{iso}(\theta_{j})$ with various models to constrain the possible
quasi-universal jet structure $\epsilon(\theta)$.  As discussed in the
introduction and in ZM, two plausible models for the jet structure
$\epsilon(\theta)$ are a power-law and a Gaussian. We describe each in
turn below.
  
 \subsection{Power-law}
   One possible model for the structure of a  universal jet is
   a power-law: 
   \begin{equation}
   \epsilon(\theta) = \epsilon_{o}(\theta/\theta_*)^{-k},
   \end{equation}
    for $\theta > \theta_{c}$, where
   $\theta_{c}$ is some cutoff to prevent divergence (see also ZM).  
    The solid line in Figure 1 shows the 
best power-law fit to $E_{iso}(\theta_{j}) \propto \epsilon(\theta)$.
The best fit gives an index $k$ of $-1.9 \pm .1$. We emphasize
that the reduced $\chi^{2}$ of this fit is unacceptably high ($\sim 8$)
because of the intrinsic scatter in the data (we will return to this point when we 
compare Gaussian models to the data).  Nonetheless, a simple power-law 
structure 
describes the general trend of the data well.  Of course this result
is not surprising since the Frail et al. (2001) and BFK result shows that
 $E_{iso}\theta^{2} \sim$ constant (or $E_{iso} \propto \theta^{-2}$).  
Although this model appears to describe the data, it is worth
investigating what other types of jet structure may produce the BFK data.  
In particular, Granot \& Kumar (2003) ran simulations which showed that a
power-law jet structure cannot reproduce the afterglow light curves -
there is an unobserved flattening both right before the jet break and also 
at late times (after the jet break).
A Gaussian, model,
on the other hand, does reproduce the features observed in the afterglow
lightcurves (Kumar \& Granot, 2003).  We explore this model in the next
section.
  
  \subsection{Gaussian} A Gaussian model, 
  \begin{equation}
  \epsilon(\theta) = \epsilon_{o} e^{-\theta^{2}/2\theta_{o}^{2}}
\end{equation}
 is also a physically reasonable suggestion for the structure of a
universal jet.  In this model, the energy peaks at $\theta = 0$
(i.e. down the center of the jet axis) and $\theta_{o}$ can be considered
a characteristic width of the jet.  The dotted line in Figure 1 shows the best fit
Gaussian jet structure, with a characteristic jet width $\theta_{o}$ of 13
degrees.  By eye, one can see that this is an unacceptable description of the data
and indeed this is confirmed formally (with reduced $\chi^{2} \sim 20$).
[We note that when a Gaussian + power-law model is tried, the Gaussian
component is made negligible in the fit routine, consistent with the above
results.]
 
  However, we point out that it is unphysical to expect that a jet's
energy structure (as a function
  of angle from the axis) will be exactly the same from GRB to GRB.  A
much more natural and physical
  model is to allow a {\em quasi-universal} configuration - i.e. to allow
some scatter in the parameters
  of each of these models.  It is possible, then, that more general
configurations (other than
  a simple power-law) can better describe the data.
  We explore this possibility in the following section.
  
   \subsection{Varying the Model Parameters}
     We investigate how the Gaussian and power-law models improve in describing the data when we 
allow the parameters (such as the power-law index or mean of the Gaussian 
distribution) to vary. We put limits on how much the parameters in each model much vary to reproduce
the scatter in the observed data.  As we will show below and as is consistent with our results
in the previous section, a power-law model does a good job of describing the data and requires only small
intrinsic scatter in the model parameters.  However, the Gaussian model can also reproduce the
observed data well; futhermore, in most cases this model predicts a sharp cutoff in the $E_{iso}-\theta_{j}$ plane 
at large $\theta_{j}$ as well as a turnover at very low values of 
$\theta_{j}$, the latter of which may already have been observed.  We discuss this in more
detail below.

   \subsubsection{Method}
     To test this scenario, we allow one parameter (or two when noted)
      of each model to vary according to either a log-normal or normal distribution.  For
example we can allow the normalization of the Gaussian model $\epsilon_{o}$ to vary as
a log-normal distribution rather than be a single fixed value. 
 For each value of $\theta$,\footnote{Our $\theta$ values are taken from
the BFK data directly.  However, we find qualitatively and quantitatively
similar results if we draw our $\theta$ values from, say, a uniform
distribution in linear space.} we draw the normalization $\epsilon_{o}$ 
(in this example) from a log-normal distribution with some mean and 
scatter, and keep the mean of the jet width $\theta_{o}$ fixed to some 
value.  We then compute the energy using this normalization and 
$\theta_{o}$ according to equation 2.  Because the normalization is drawn 
from a distribution with some scatter, this will produce some scatter in 
the energy as a function of angle.  The point is to find out how {\em 
much} scatter is needed in each parameter of the models to reproduce the 
data.
  Besides requiring the data to qualitatively {\em look} similar, by
reproducing the energy range and scatter of the BFK data, for the Gaussian
jet model we would like the simulated data to be fit by a power-law with
an index consistent with the fit to the BFK data ($\sim -2$), with a
$\chi^{2} \sim 4 - 10$ (similar to that of the data).  Finally, we run a
Kolmogorov-Smirnov (KS) test to check whether the two distributions (data
vs. simulations) can be considered statistically similar.

  We comment that for all parameters but the index $k$ in the power-law model, the log-normal distribution is the
  probably the most physically reasonable distribution for adding dispersion to the parameters of each model. It 
  accurately reflects the dynamic range of variation seen in the data and avoids unphysical negative values for 
  the energy that can arise when a simple normal distribution is used for 
the parameter variations.  Also, if GRBs
  from progenitors over a wide mass range, e.g. $10-300 M_{\odot}$, a
log-normal distribution is more natural to describe the dispersion in the 
physical parameters.
  If, however, they arise from progenitors of a very narrow mass range, a normal distribution for parameter variations may be
  appropriate.  Although (except for the power-law index $k$)
  we consider the log-normal variations to be physically more reasonable, we present the 
  results for both distributions below; our figures, however, show only 
the results of the log-normal variations (except, again, for the power-law 
index $k$).  

   \subsubsection{Results}
     \noindent {\bf 1. Power law:}\\ \indent First, we explore how much
scatter is necessary in the power-law model to reproduce the scatter of
the BFK data.
      We consider the case where the power-law index $k$ in equation 1 is
not a constant but
     derived from a
     normal distribution $\propto \ {\rm
exp}^{-(k-<k>)^{2}/2\sigma_{k}^{2}}$.  We choose a mean $<k>$ of -2 and
     explore various $\sigma_{k}$ values that can reproduce the data.  
For a
     $\theta_{*} = 1^{\circ}$ in equation 1, a scatter of $\sigma_{k} \sim
0.7$ adequately
     reproduces
     the BFK data.  This is shown by the circles in Figure 2 (the BFK data
are the square points).
         The bottom
     line is that the scatter in the power-law index need be on the
order of $\sim 0.7$ to reproduce
     the scatter in the BFK data.
      We also vary the normalization
     $\epsilon_{o}$ as a log-normal
     $\propto \epsilon_{o}^{-1}{\rm exp}^{-({\rm lg}(\epsilon_{o})-<{\rm
lg}(\epsilon_{o})>)^{2}/
     2\sigma_{{\rm lg}(\epsilon_{o})}^{2}}$. We find that a mean of $<{\rm
lg}(\epsilon_{o}/10^{50}{\rm erg})> =
     4.8$ and a dispersion $\sigma_{{\rm lg}(\epsilon_{o}/10^{50}{\rm
erg})} = 0.6$ adequately reproduces the data (in the sense of reproducing
the scatter and preserving the $k=-2$ power-law behavior)
     as shown by the circles in Figure 3.  When we vary $\epsilon_o$
according to a normal distribution $ \propto \ {\rm
exp}^{-(\epsilon_{o}-<\epsilon_{o}>)^{2}/2\sigma_{\epsilon_{o}}^{2}}$,
     we find that $<\epsilon_{o}/10^{50}{\rm erg}> = 1.6 \times 10^{4}$
and
     $\sigma_{\epsilon_{o}} = <\epsilon_{o}>$ does an adequate job of
     reproducing the scatter.  In
all cases, we have taken $\theta_{*}$ in
     equation 1 equal to $1^{\circ}$.

     \noindent {\bf 2. Gaussian:}\\  
     In this case, we tried varying each of the parameters, $\epsilon_{o}$ 
and $\theta_{o}$ of the Gaussian
model in equation
     2, according to a log-normal and a normal distribution.  For example, as described below, we let
the dispersion or
     characteristic width of the jet $\theta_{o}$ be derived
     from  a log-normal function with a mean $<{\rm lg}(\theta_{o})>$ and 
     standard deviation $\sigma_{{\rm lg}(\theta_{o})}$,
     while $\epsilon_{o}$ is fixed to some constant value.
    In addition, we try a model in which the
     normalization $\epsilon_{o}$ is taken from a log-normal distribution
and then $\theta_{o}$ varies in a
     correlated way according to the equation $\epsilon_{o}\theta_{o}^{2}
\sim$ constant.  This is motivated by the fact that the Frail et al.
(2001) and BFK data imply that the total GRB energy is approximately
constant (in a uniform jet configuration); in the quasi-universal Gaussian
model, this total energy is given by $\sim \epsilon_{o}\theta_{o}^{2}$ 
(ZM).
      A summary of our results are as
     follows:
     
     \begin{itemize}
     \item{Varying the characteristic jet width, $\theta_{o}$: We try a model in which $\theta_{o}$ in equation
     2 varies as a log-normal with mean  $<{\rm lg}(\theta_{o})>$ and 
     standard deviation $\sigma_{{\rm lg}(\theta_{o})}$.  In general, we
find that $<{\rm lg}(\theta_{o}/1^{\circ})>$ in the
     range $0.7-0.8$ and $\sigma_{{\rm lg}(\theta_{o})}$ in the range 
0.3-0.5 does the best job reproducing the
     data.  A higher mean tends to create an unobserved flattening at low values of $\theta_{j}$, while a lower
     mean causes the simulated data to be too steep (with power law index
much steeper than $-2$).  Figure 4 shows the simulated data vs. actual
data
      in the BFK range
       for $<{\rm lg}(\theta_{o}/1^{\circ})> = 0.8$ and $\sigma_{{\rm
lg}(\theta_{o})} = 0.4$.  The simulated
       data shown in Figure 4 are fit with a power-law of index $-1.8 \pm
0.2$ in the range of the BFK data; this is consistent with the index we
obtained when fitting a simple-power-law to the BFK data.  The fit gives a
reduced $\chi^{2} = 4.0$, which - although formally unacceptable - is
consistent with the value of $\chi^{2}$ obtained when we fit a power-law
directly to the BFK data.  Hence, we consider this
     similar value of $k$ and $\chi^{2}$ to indicate that we are
accurately reproducing the slope and scatter of the
     data. Finally, when we perform a KS test on the simulated and
observed data, we find that the
     probability that these data are derived from the same distribution is
$69$ \%.  A very low value of this
     probability $<<1$\% would indicate with statistical robustness that
these data are from different distributions.
     Hence, we consider this model to adequately reproduce the data.  
     
     Using a normal distribution to produce variation in
     $\theta_{o}$, with  average value $<\theta_{o}>$ and a standard deviation
     $\sigma_{\theta_{o}}$,  
      we find that  values $<\theta_{o}> \approx 7-10^{\circ}$ and 
$\sigma_{\theta_{o}} \approx
       2-4^{\circ}$ do the best job in reproducing the data.  Our best fit 
was 
       for $<\theta_{o}> = 9^{\circ}$ and $\sigma_{\theta_{o}} = 
4^{\circ}$. The simulated 
       data in this case are fit with a power-law of index $-2.0 \pm 
0.2$ in the range of the BFK data; this is consistent with the index we 
obtained when fitting a simple-power-law to the BFK data.  The fit gives a  
reduced   $\chi^{2} = 6.4$.  A KS test gives a probability of $15$ \% 
 that these data are derived from the same distribution.  In general, this
particular model is quite unstable (for small variations in $\theta_{o}$, 
we can get
huge variations in the energy) and reproduces the data only in special
cases.
  }

       \item{Varying the normalization, $\epsilon_{o}$:  
      We vary the normalization $\epsilon_{o}$ as
       according to a log-normal distribution with mean $<{\rm
lg}(\epsilon_{o})>$
        and standard deviation $\sigma_{{\rm lg}(\epsilon_{o})}$.  For a
$\theta_{o} = 10^{\circ}$,
	  $<{\rm lg}(\epsilon_{o}/10^{50}{\rm erg})>$ in the range
$3.4-3.7$ and a
        $\sigma_{{\rm lg}(\epsilon_{o}/10^{50}{\rm erg})}$ in the range
$0.5-0.7$ describes the data well.
	Figure 5 shows the simulated data for $<{\rm
lg}(\epsilon_{o}/10^{50}{\rm erg})>=3.6$ and
	$\sigma_{{\rm lg}(\epsilon_{o}/10^{50}{\rm erg})}=0.7$
	 The simulated
       data shown in Figure 5
       are fit with a power-law of index $-1.5 \pm 0.3$ and $\chi^{2} =
5.6$ in the range of the
     BFK data.  The formal power-law fit is marginally consistent with 
the fit to the BFK data, but a KS test performed on the simulated and 
observed data sets gives a probability of $99$ \%
     that the two data sets are derived from the same parent 
distribution.\footnote{Again, we note that the KS diagnostic is intended 
to check whether two distributions can be considered statistically {\em 
different}.  A value $<< 1$ \% would indicate such a result.  A higher
value (like $99$ \%) does not necessarily indicate the model is any better 
than a lower value (like $50$ \%).  All of our KS probabilities show that 
it is {\em not} the case that the simulated and observed data are derived 
from different distributions. }
       
       When we vary the normalization $\epsilon_{o}$ according to
       a normal distribution with mean $<\epsilon_{o}>$ and standard
deviation $\sigma_{\epsilon_{o}}$,
       we find that for a $\theta_{o} = 10^{\circ}$,
	  $<\epsilon_{o}/10^{50}{\rm erg}> = 1.3 \times 10^{3} $ and
        $\sigma_{\epsilon_{o}} = 1.7 \times 10^{3} $ describes the data 
well.  Note that the values of the mean and dispersion cause some deviates to be negative.
	In this case, we take the absolute value of all deviates.  
     The data here are fit with a power-law of index $-2.0 \pm 0.2$ and 
$\chi^{2} = 7.4$ in the range of the
     BFK data.  A KS test performed on the simulated and observed data
sets gives a probability of $49$ \%
     that the two data sets are derived from the same parent distribution.}   
       
       \item{ Varying $\theta_{o}$ and $\epsilon_{o}$ in a correlated way:  Finally, we vary
        $\theta_{o}$ and $\epsilon_{o}$ such that the total energy 
$\epsilon_{o}\theta_{o}^{2} = constant$, again motivated by the result of 
Frail et al. (2001) and BFK that the total GRB energy is approximately 
constant.  To do this,
	we take $\epsilon_{o}$  from a log-normal distribution as described in the previous section 
	and solve for $\theta_{o}$ from the above relationship.  
	We find the observed data is reproduced when $<{\rm
lg}(\epsilon_{o}/10^{50}{\rm erg})>=3.4$ and
	$\sigma_{{\rm lg}(\epsilon_{o}/10^{50}{\rm erg})}=0.6$.  This
results in a mean jet
	width of $20^{\circ}$. The simulated 
       data  shown in Figure 6 
       are fit with a power-law of index $-1.8 \pm 0.2$ and $\chi^{2} \sim 
6.6$ in the range of the
     BFK data.  A KS test performed on the simulated and observed data
sets gives a probability of $63$ \%
     that the two distributions are the same. 

	For a normal variation in $\epsilon_{o}$, 
	we find the observed data is reproduced when
$<\epsilon_{o}/10^{50}{\rm erg}> = 1.3 \times 10^{3} $ and
        $\sigma_{\epsilon_{o}/10^{50}{\rm erg}} = 2.0 \times 10^{3}$. This
gives a mean jet width of
        about $10^{\circ}$.  These simulated
       data 
       are fit with a power-law of index $-1.7 \pm 0.2$ and $\chi^{2}
= 4.4$ in the range of the
     BFK data.  A KS test performed on the simulated and observed data
sets gives a probability of $21 $ \%
     that the two distributions are the same.
      }
     \end{itemize}
     
     The bottom line is that the data can be adequately reproduced with a
Gaussian model for the jet structure, given some variation in the models
parameters (we note that in our simulations, we have assumed that the
viewing angle is the agent that determines the jet break time).
 Each model above was able to reproduce the power-law behavior of the BFK
data, the scatter (reflected by the value of $\chi^{2}$ of the power-law
fit), and a KS test probability that shows that the simulated data are not
inconsistent with being derived from the same parent distribution as the
observed data.  It appears, however, that the log-normal variations in
$\epsilon_{o}$ and the correlated $\epsilon_{o}-\theta_{o}$ variations do
the best job in reproducing the data (see Figures 5 and 6).  The model in
which $\theta_{o}$ varies in general exhibits too much curvature and is
less stable to the degree of variation.  Although the correlated
$\epsilon_{o}-\theta_{o}$ variations do a good job of reprocing the data,
the additional constraint that $\epsilon_{o}\theta_{o}^{2} = constant$ 
may
be considered an unattractive feature of this model.  However, we do point
out that this constraint is motivated by the observations (Frail et al.,
2001, BFK) and may in fact be a physical feature of GRBs.
    
     An interesting aspect of the Gaussian model is that the $k=-2$
power-law behavior is not
     extended below values of $\theta_{j}$ $\la 2^{\circ}$ and in some
cases above $\theta_{j} \ga 30^{\circ}$.  Figure 7 shows simulations for
the correlated $\epsilon_{o}-\theta_{o}$ variations, 
extendend above and
below the range of the BFK data.  A turnover is clearly seen at low values
of $\theta_{j}$.  For the case of either $\epsilon_o$ or $\theta_{o}$
varying individually, or for the case of the correlated
$\epsilon_{o}-\theta_{o}$ with a {\em low value for the average jet width}
$\la 5^{\circ}$, we also find a cutoff at high values of $\theta_{j} \ga
30^{\circ}$.  We note that Figure 7 uses the same parameters as in Figure 
6, where the average jet width is $20^{\circ}$. This average jet width 
can vary and this will affect exactly where the low angle turnover and 
high angle cutoff appear.
       Nonetheless, if this model were correct, then we should
       not see many more bursts with $\theta_j \la 1^{\circ}$, and possibly also
       $\theta_j \ga 30^{\circ}$.  These predictions may be tested with future 
observations of
       GRBs, and in fact the low $\theta_{j}$ prediction may have already
       been seen. Interestingly, BFK discuss several outliers in their sample that
       appear to be ``fast-fading'' (possibly because the jet break occured at very early times, implying a
       very small value for the jet opening angle $\theta_{j}$) {\em and} sub-luminous relative to the rest
       of the sample.  This is consistent with the trend predicted by the
Gaussian jet structure at small values
       of $\theta_{j}$.   

\section{Luminosity Functions}
  \subsection{Predictions} One of the advantages to the quasi-universal jet configuration, is that
  it makes definite predictions for the observed luminosity function (LF) of GRBs.  For example, as discussed in ZM,
  the isotropic luminosity function $N(\epsilon)$ at a given redshift
  can be determined from the burst angular distribution $N(\theta)d\theta \propto
  {\rm sin}\theta d\theta \propto \theta d\theta$.  In the power-law model, $\theta \propto \epsilon^{-1/k}$, so that
  $N(\epsilon) \propto \theta d\theta/d\epsilon \propto \epsilon^{(-1-2/k)}$.  For a Gaussian model, $N(\epsilon) \propto
  \epsilon^{-1}$.  These predictions are derived in the case of no variations in the model parameters.
    We extend this analysis by simulating luminosity functions, allowing parameters
  to vary in our power-law and Gaussian models.   
Using the best fit parameters and variation schemes from
the previous analysis, we simulate 500,000 GRBs for each jet structure  with 
the distribution of observing angle following 
$N(\theta_{v})d\theta_{v}\propto\sin{\theta_{v}}d\theta_{v}$, and where the range
 of the observing angle is from $0^{\circ}$ to $90^{\circ}$.
Again, these simulations hold for luminosity functions at a given redshift.

\subsection{Power-law Jet Structure}
We first test our methods by simulating the GRB LF in the case where
$\epsilon(\theta)$ is a simple power-law (with no parameter variations, and our best fit
index of $k=-1.9$).
As expected we obtain a slope of the LF of index $\approx -2$ (solid line in
Figure~\ref{fig:lf1}), consistent with the
analytical expression of ZM and given above in \S 4.1.  The minimum luminosity
of $\sim 10^{51}$ erg is due to the upper boundary of the observer angle $\theta_{v} = 90^{\circ}$.
We then simulate the cases where the power-law index $k$ varies according to a normal
distribution, and 
$\epsilon_{o}$ varies according to both normal and log-normal distributions.
The results are shown as dotted, dashed, and dash-dotted lines in 
Figure~\ref{fig:lf1}, respectively.
The luminosity functions obtained when we allow parameters to vary are similar to 
the simple power-law jet at the high luminosity end, being well described by a power-law of index $-2$ for 
more than three order of magnitudes.
At the low luminosity end, however,  the LFs tend to flatten.
This flattening is a result of the low-luminosity cutoff present in the simple
power-law case.  For example, above the cutoff, when we allow parameters to vary there
is a compensation between lower and higher luminosities that preserves
the overall slope of the luminosity function.  However,  variations in the parameters near the cutoff cause
some GRB luminosities to be  pushed below the minimum luminosity, without compensation from
GRBs of lower luminosity.   
 
\subsection{Gaussian Jet Structure}
Again, we first simulate the luminosity function of a Gaussian jet, with
fixed normalization, $\epsilon_{o}$, and characteristic jet width, $\theta_{o}$.
The result is a powerlaw of slope -1, shown by the solid line in 
Figure~\ref{fig:lf2}, and consistent with the analystical expression in \S 4.1.
Note that this luminosity function has a maximum luminosity corresponding to an 
observer angle of $0^{\circ}$.
We then vary the characteristic angle, $\theta_{o}$, of the Gaussian jet 
according to a log-normal distribution.
The simulated luminosity function is very similar to the luminosity 
function obtained from constant Gaussian jet model; it has a power-law index 
of $-1$ and an upper
bound, shown as dotted line in Figure~\ref{fig:lf2}.
Next, we consider the case where $\epsilon_{o}$ varies as log-normal.
The result is shown by the dashed line in Figure~\ref{fig:lf2}.
The luminosity function resembles the case of constant $\epsilon_o$ and $\theta_o$
on the low luminosity 
end, but steepens at high luminosities.
This steepening is due to the same type of effect that caused the flattening at the
low luminosity end in
the power-law jet case (see \S 4.2 above) - variations in the parameters near the
maximum luminosity cause some GRB luminosities to be pushed above the maximum, without
compensation from GRBs of higher luminosity.
We also simulate the luminosity function in the case where both
$\epsilon_{o}$ and $\theta_{o}$ vary, but preserve the relation $\epsilon_{o}\theta_{o}^{2} = 
constant$ (see \S 3).
The luminosity function is shown as dash-dotted line in 
Figure~\ref{fig:lf2}.
The LF in this case has a slightly flatter power-law index at low luminosities,
with a clear break at $L \sim 10^{52}$ erg and a sharp steepening after that.
This break occurs because an anti-correlated $\epsilon_{o}-\theta_{o}$ distribution causes an enhanced
pile-up near ($<{\rm lg}(\epsilon_o)>,<{\rm lg}(\theta_o)>$) in the $\epsilon-\theta$ plane, leading to
a flattening of the LF at the corresponding luminosity.
Finally, we simulate LFs from Gaussian jets where the parameters vary according to
a normal distribution; the LFs are shown in Figure~\ref{fig:lf3}.
The qualitative behavior is similar to the log-normal variation case, although these
LFs are steeper at the high luminosity end over a smaller dynamic range.
 
The luminosity functions in both the power-law and Gaussian 
jet structures have a definite change of power-law index when variations in the model
parameters are included.  
Power-law jets produce LFs with a slope of $-2$, but with a dramatic
flattening at luminosities below about $10^{51}$ erg.
Gaussian jets produce LFs with a slope of $\sim -1$, although varying 
$\epsilon_{o}$, and $\theta_{o}$ and $\epsilon_{o}$ in an anti-correlated way, causes the LF
to steepen at high luminosities.  In particular, the anti-correlated $\epsilon_{o}-\theta_{o}$
case with log-normal parameter variations produces a clear break at $\la 10^{52}$ erg,
 above which there is a steepening with
an index of $\sim -2$.

  \subsection{Comparing with observed data}
    We would like to compare the predictions for the GRB luminosity
function from the quasi-universal jet model
    of GRBs, with the observed GRB luminosity function.  Unfortunately, we are not yet at the point where we can
    directly measure the GRB LF with any statistical certainty. This is because there are still on a handful
    $(\sim 33)$ of GRBs with measured redshifts, and determination of the
LF requires large numbers (at least 100) of luminosities
    from a uniform sample (and a sample with complete and detailed knowledge of the selection effects). 
    
     Usually, the GRB LF is estimated by utilizing
    a large sample of GRBs with a measured flux distribution $n(f)$,
     {\em assuming} an underlying density distribution, $\rho(z)$, 
cosmological model, and then
    computing the luminosity function $N(L)$ through the equation $n(f)df 
= N(L)\rho(z)/(1+z) (dV/dz) dL dz$.  Note that
    this assumes independence of the variables $L$ and $z$, or - in other words - that the luminosity function
    does not evolve with redshift.  Several studies have taken this approach to estimate the GRB LF.  For example,
    Stern et al. (2003) - {\em assuming} a steep decline in the GRB source population above a redshift of 1.5 - 
    find $N(L) \propto
    L^{-1.4}$ up to $L\sim 3 \times 10^{51}$ erg and then sharply declines after that.  Schmidt (2002) - using various models for
    the GRB rate density based on the estimates for the global star 
formation rate - find that the GRB luminosity function
    can be described by a broken power-law with an index $\sim -1$ on the low luminosity end ($< 3 \times 10^{51}$ erg) and
    sharply declines after that, consistent with the Stern et al. results.  
[All luminosities quoted are the
    isotropic equivalent luminosities.]   Although these results are fairly uncertain because they rely on the assumption
    of the underlying GRB rate density, they appear to be marginally consistent with predictions from the Gaussian jet structure.
    The power-law universal jet may also be consistent with these results if the data is fit
    with a single power-law such that the steepening of the LF above some break causes an overall steepening of the LF index,
     closer to
    $-2$.   [We note that recently, some
    attempts have been made to constrain the GRB LF based on the small sample of bursts with direct spectroscopic redshifts.
      For
    example, van Putten \& Regimbau (2003)
     find that a log-normal description of the LF does a suitable job of describing the existing
    data.]

    Although spectroscopic redshifts have only been obtained to a number
of GRBs too small to determine a luminosity function
    directly, in recent years there has been evidence of so-called ``luminosity indicators'' (Fenimore \& Ramirez-Ruiz, 2000,
    Reichart et al., 2001, Norris et al. 2001) from which luminosities of GRBs can be obtained.  For example, Fenimore
    \& Ramirez-Ruiz (2000) found a correlation between GRB luminosity and the light curve variability (an estimate of
    the ``spikiness'' of the light curve) based on a small sample of
bursts with measured redshifts.  From this, they were
    able to obtain luminosities and redshifts to 220 GRBs.  Because they 
used BATSE gamma-ray
    data and chose a flux limit to their sample, the selection effects were very well defined and therefore a luminosity
    function can be estimated.  Fenimore \& Ramirez-Ruiz assumed independence between $L$ and $z$ in their sample and found a 
    luminosity function with a power-law index of approximately $-2$, consistent with the power-law quasi-universal jet
    configuration.  A more detailed analysis of this sample, however, shows that $L$ and $z$ are in fact 
    correlated (Lloyd-Ronning,
    Fryer, \& Ramirez-Ruiz, 2002, Yonetoku et al. 2003, Graziani et al. 2003); this correlation has to 
be accounted for to correctly determine a LF from the
    data.  Lloyd-Ronning et al. did this by defining a variable $L' =
L/\lambda(z)$, where $\lambda(z)$ parameterizes the
    correlation between $L$ and $z$.  They found a single power-law 
fit gives  $N(L') \propto
L'^{-2.2}$, while a broken power-law fit gives $N(L') \propto
    L'^{-1.5}$ for $L'<L'_o$ and $N(L') \propto
    L'^{-3.3}$ for $L'>L'_o$, where $L'_o = 5 \times 10^{51}$ erg. Again, there is no underlying assumption about
    the density distribution of GRBs in this analysis and the correlation between $L$ and $z$ present in the
    data has been accounted for. [We mention that in the context of
    the quasi-universal jet model, luminosity evolution 
    could imply the overall normalization $\epsilon_{o}$ evolves with redshift, or that the characteristic
    opening of the jet $\theta_{o}$ evolves.]  
    The single power-law result appears to be more consistent with the
power-law quasi-universal
    jet structure, while the
    broken power-law result is better described by the (at least qualitative) predictions of the Gaussian case.
    Finally, Norris (2002) found  - using the luminosities and redshifts 
obtained from the luminosity-lag relation, another luminosity indicator (Norris et
al, 2001) - that the GRB LF  scales as $L^{-1}$ for low 
luminosities
    and $L^{-1.8}$ for high luminosities.  This is very similar to the 
results we obtain for the LF from
    a Gaussian jet when $\epsilon_o$ and $\theta_o$ vary in an anti-correlated way.
      
    Although some of the above results suggest the plausibility of a Gaussian jet structure for GRBs,
    the predictions of the quasi-universal jet model  GRB LFs are better tested when more GRBs with
    redshifts are obtained and a direct GRB LF can be measured.  This is a 
primary goal of the Swift satellite, to be launched in June of 2004. 
 
\section{Summary and Discussion}
   In this paper, we have tested two possible models for a quasi-universal jet structure to a GRB.  In particular, we attempt to 
   reproduce the data of BFK who showed an anti-correlation between the isotropic emitted
   energy $E_{iso}$ and the jet opening angle, $\theta_{j}$ derived in a
uniform jet model, such that $E_{iso} \propto \theta_{j}^{-2}$.
   In the quasi-universal jet model (Rossi et al., 2002, ZM), in which the jet opening angle and total
   emitted energy are approximately the same for all GRBs, but the energy varies as a function of angle from the jet axis {\em
   within} a GRB, this correlation is a reflection of the energy profile
$\epsilon(\theta)$ as a function of angle from the jet axis.
   Motivated by ZM, who suggested that this profile may take on different functional forms, we have tested how Gaussian and 
   power-law models for $\epsilon(\theta)$ reproduce the BFK data. 
 In particular, we have allowed for realistic scatter in the parameters of 
 each model (letting each parameter vary as a either a log-normal or normal
 distribution with some mean and
standard deviation).
  
  We find that both the power-law and Gaussian jet structures can reproduce
the data quite well, with only minimal scatter required in the model
parameters  - particularly when the parameters vary according
to the more physically reasonable log-normal distribution.  The strengths of the power-law model
are its simplicity and ease in reproducing the observed data. However, as discussed in the 
introduction and throughout the text, there are several reasons to consider other jet
configurations, and in particular the Gaussian model appears to adequately describe the observations.
Furthermore,
 The 
Gaussian model predicts a sharp turnover
in the $E_{iso}-\theta_{j}$ plane  for low (and in some cases high) values of $\theta_{j}$.  
This turnover at low jet angles
has possibly been observed by BFK, who found a few ``sub-luminous'' bursts (relative
to the rest of their sample), with steeply declining light curves which could indicate a very small opening
angle $\theta_{j}$.  These ``outliers'' are consistent with the trend predicted by the Gaussian quasi-universal
jet configuration at low $\theta_{j}$.
 Furthermore, these outlier bursts challenge the BFK conclusion that
the energy resevoir in GRBs is approximately constant, when taken in the context of
the uniform jet model.  However,  the
measured $\theta_{j}$ is - in the context of the quasi-universal jet paradigm - the {\em viewing} angle
and not the characteristic width of the jet $\theta_{o}$, which could be larger. This means that
the total energy within the GRB could still be standard, if the results 
are considered  in the quasi-universal jet picture with a {\em Gaussian} jet structure. 
 We also find that the luminosity function predicted by the Gaussian 
model, including variations in the model parameters, appears to be consistent with past studies of the  
luminosity GRB LF, although these predictions are better tested when the GRB LF can be directly measured.
 
   Perna et al. (2003) showed that a universal jet model with a power-law
   structure (of index of $-2$) predicts a distribution of break times in the afterglow light
   curve that is consistent with what is observed.  It 
would be interesting to explore their
   analysis in the context of the Gaussian model, which is proving to be a 
viable model for the quasi-unviersal jet structure of GRBs. (Given our 
results above, we suspect that
   a Gaussian model may give qualitatively similar results as the
power-law.)  We note that Lamb et al. (2003) have recently pointed out
that the quasi-universal jet model fails to reproduce the large dynamic
range of the observed relationship between isotropic energy and spectral
peak energy (see, e.g., Amati et al., 2002), which spans not only the
"classic" GRB energies (i.e., those of BATSE, from $\sim 50$ keV to $1$
MeV), but also includes so-called X-ray flashes, which have spectral peak
energies down to a few keV.  Their conclusions are in the framework of a
power-law structure for the quasi-universal jet, and such an investigation
in the framework of the Gaussian model is underway (Zhang et al., in
prep).
 Furthermore, we comment that the
realistic jet struture might not be strictly power-law, Gaussian, nor their
simple superpositions; in fact, recent numerical simulations 
(Zhang, Woosley, \& MacFadyen, 2003)
indicate a double Gaussian structure for the jet may provide the best description (where
one Gaussian is used for the core of the jet and one for the wings). The bottom 
line is now that we can reproduce the
data with varying parameters in these simplified models, we can do it with 
more realistic models too. And such models may be able to explain
all of the observed gamma-ray burst data in the context of a quasi-universal jet configuration.

\begin{acknowledgements}
  We would like to thank Dale Frail for sending us a table of the BFK
data.  We would also like to thank the referee for insightful comments and
suggestions that improved this work.
 \end{acknowledgements}

\clearpage

\epsscale{0.6}
\begin{figure}
\plotone{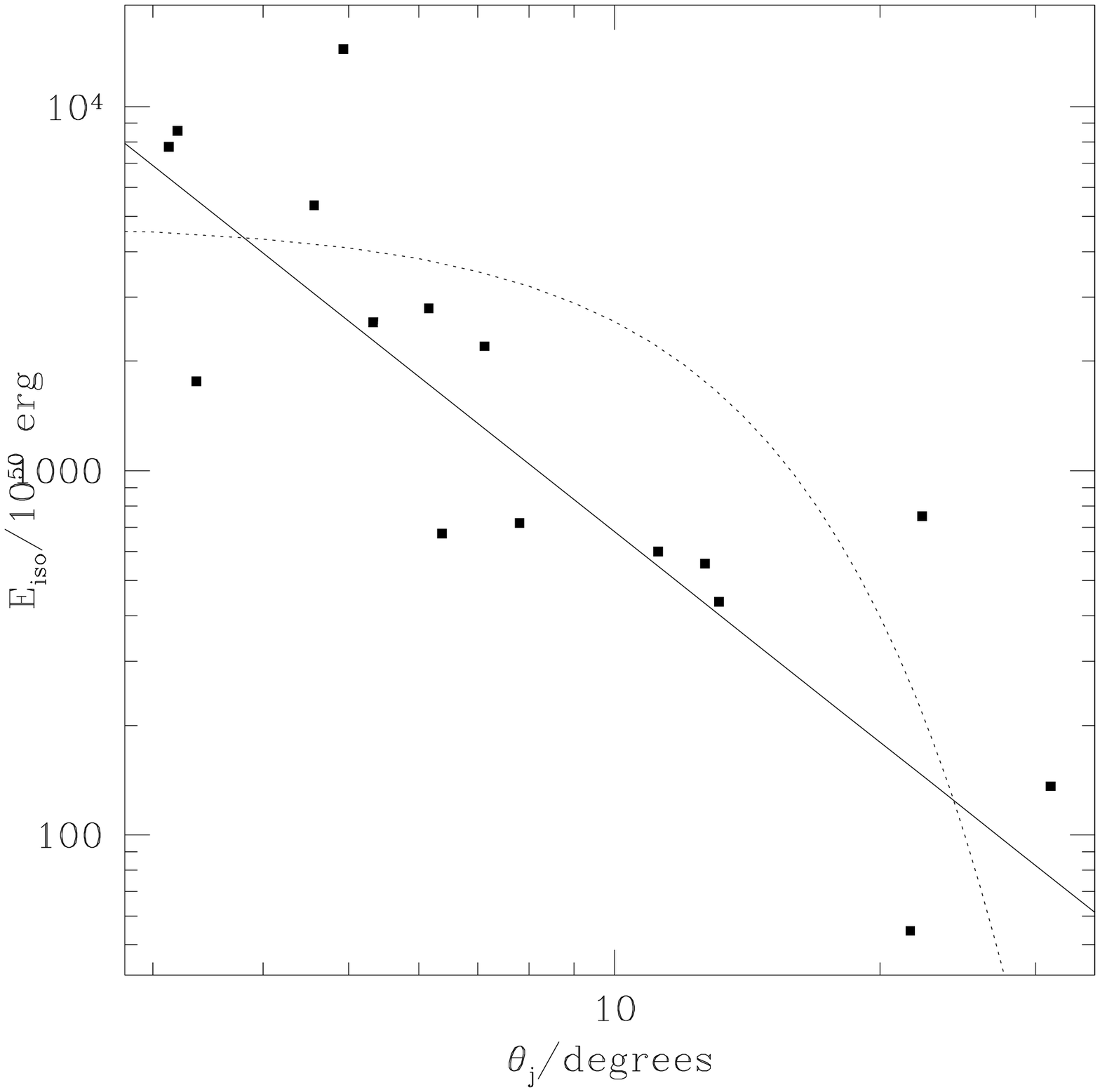}
\caption{Isotropic emitted energy vs. jet opening angle.  The square 
points are 
the data from BFK, the solid line shows a power-law fit to the data, 
while the dotted line shows a Gaussian fit. }

\plotone{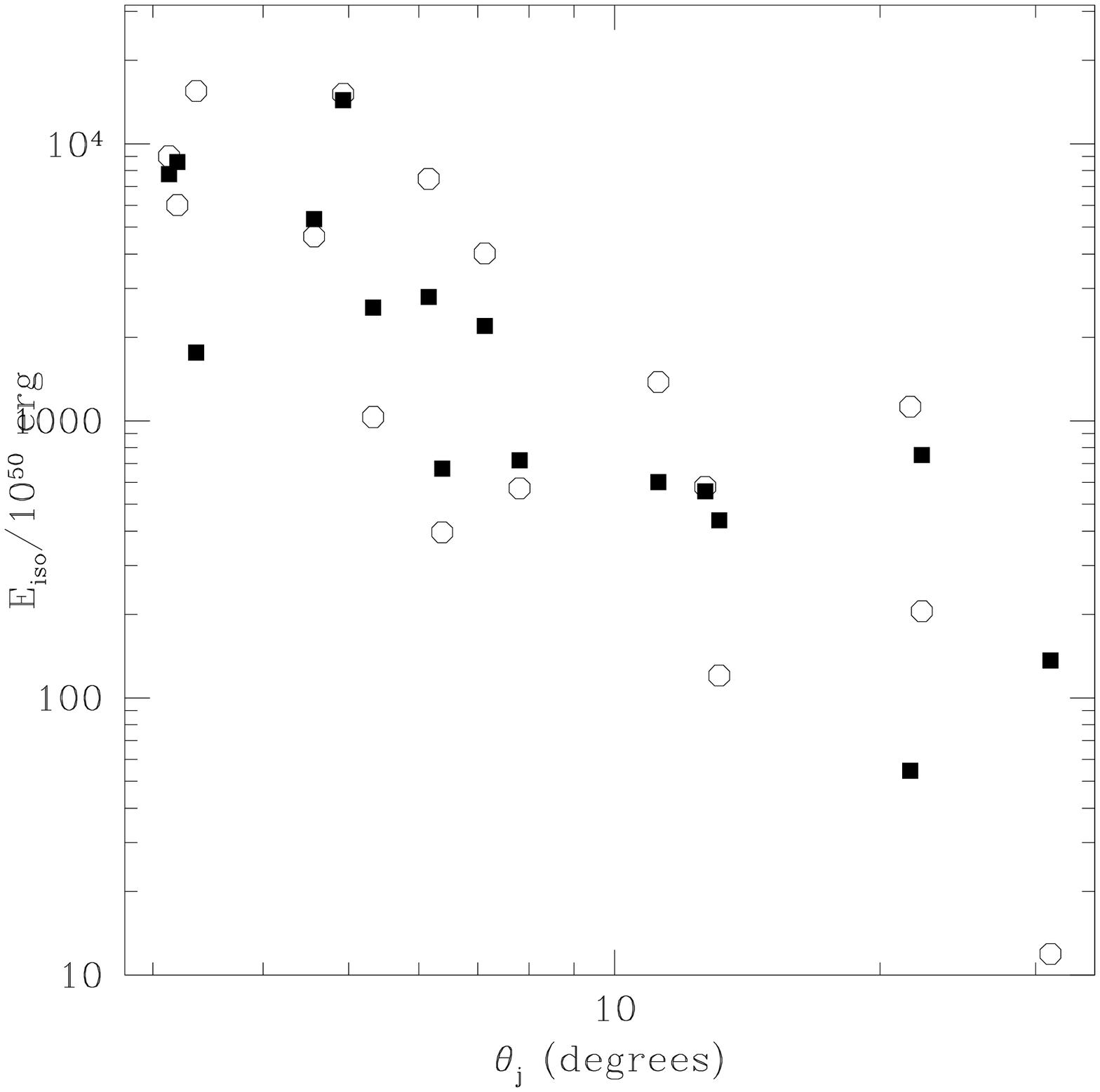}
\caption{Isotropic emitted energy vs. jet opening angle.  The square
points are the BFK data, while the circles indicate data derived from a 
power-law jet structure, with an index $k$ that varies according to a normal
distribution with
mean of $-2$ and standard deviation of $0.7$. 
} 
\end{figure}

\begin{figure} \plotone{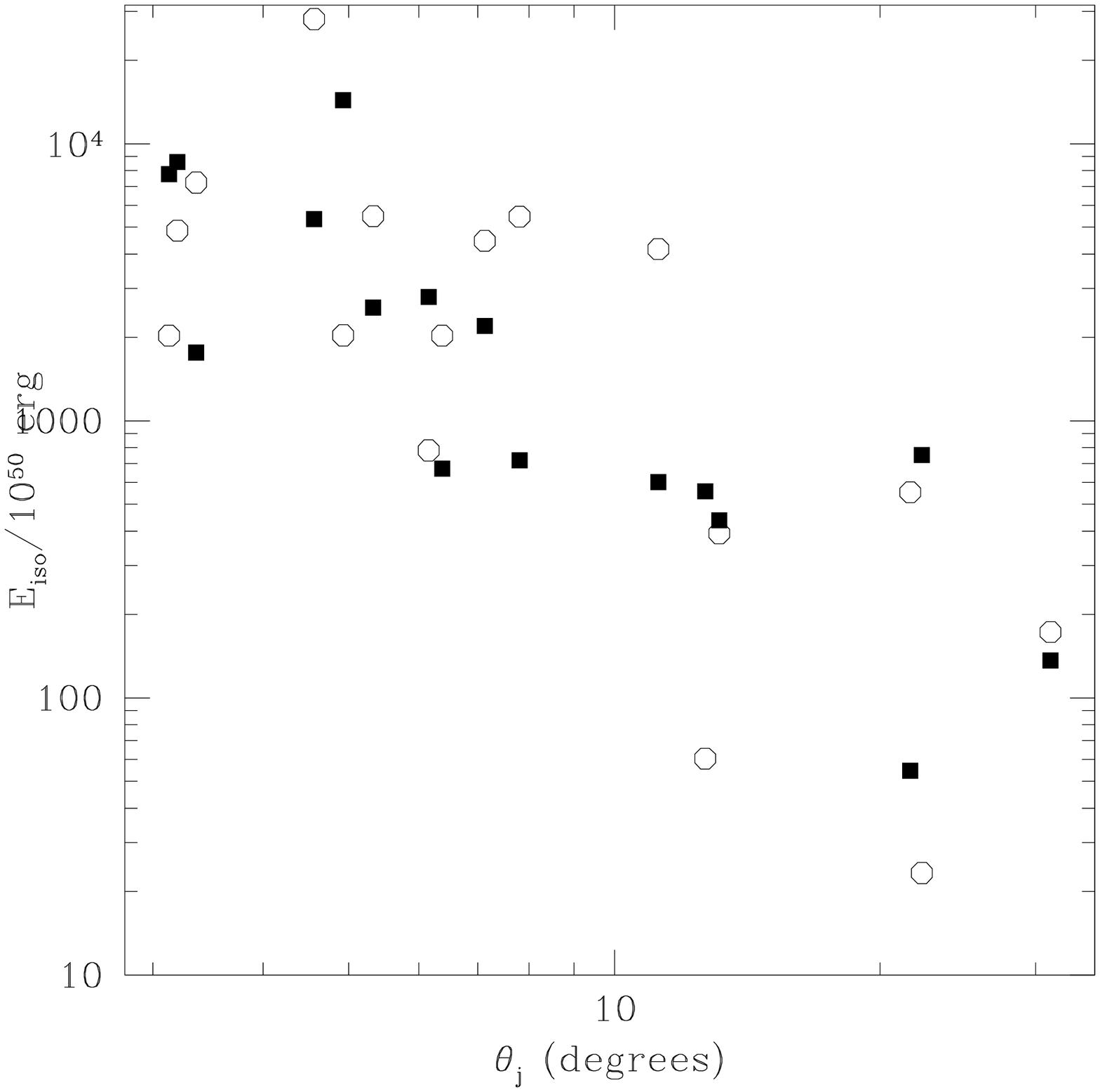}
 \caption{ Same as
Figure 2, but where the circles indicate data derived from a power-law jet
structure, where the normalization varies as a log-normal distribution,
with a mean $<{\rm lg}(\epsilon_{o}/10^{50}{\rm erg})> = 4.8$ and a
standard deviation $\sigma_{{\rm lg}(\epsilon_{o}/10^{50}{\rm erg})} =
0.6$.} 

 \plotone{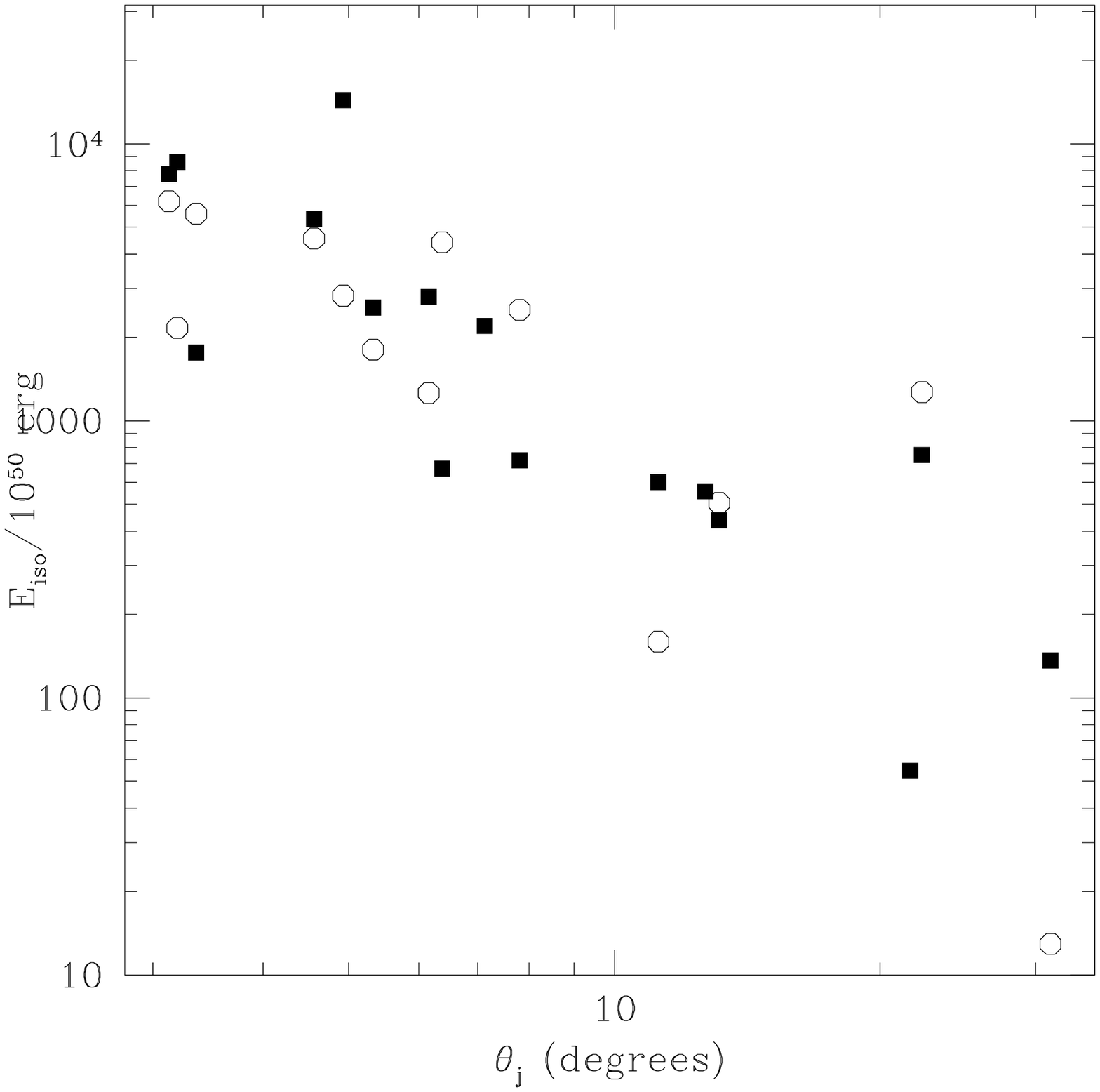} \caption{ Same as
Figure 2, but where the circles are data derived from a Gaussian jet
structure, with a characteristic width $\theta_{o}$ that varies as a
log-normal distribution with a mean $<{\rm lg}(\theta_{o}/1^{\circ})> =
0.8$ and a standard deviation $\sigma_{{\rm lg}(\theta_{o}/1^{\circ})} =
0.4$.}
\end{figure}

\begin{figure} \plotone{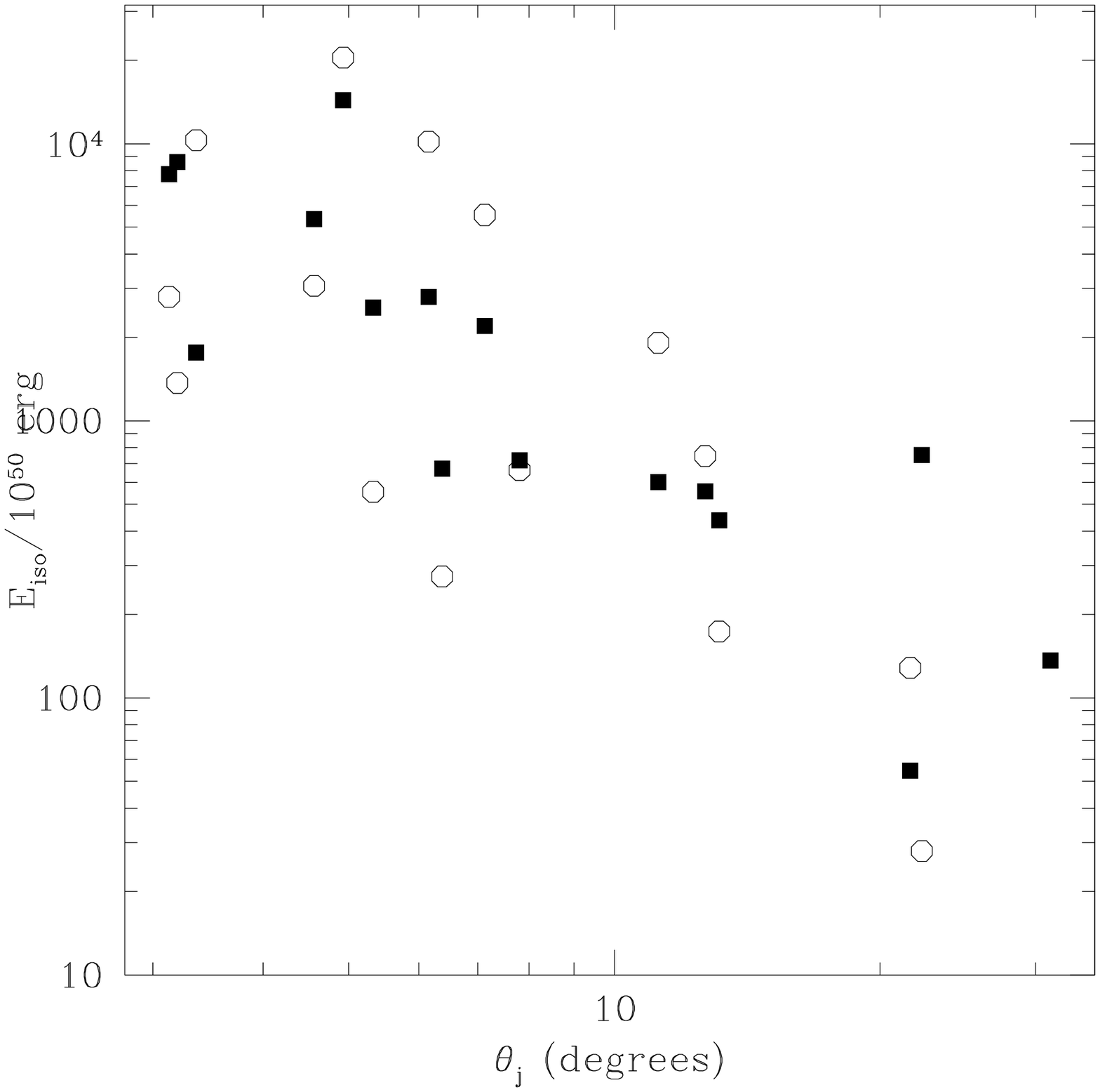} \caption{ Same as
Figure 2, but where the circles are data derived from a Gaussian jet
structure, with a characteristic width $\theta_{o} = 10^{\circ}$ and a
normalization $\epsilon_{o}$ that varies as a log-normal with a mean
$<{\rm lg}(\epsilon_{o}/10^{50}{\rm erg})> = 3.6$ and a standard deviation
$\sigma_{{\rm lg}(\epsilon_{o}/10^{50}{\rm erg})} = 0.7$.}

 \plotone{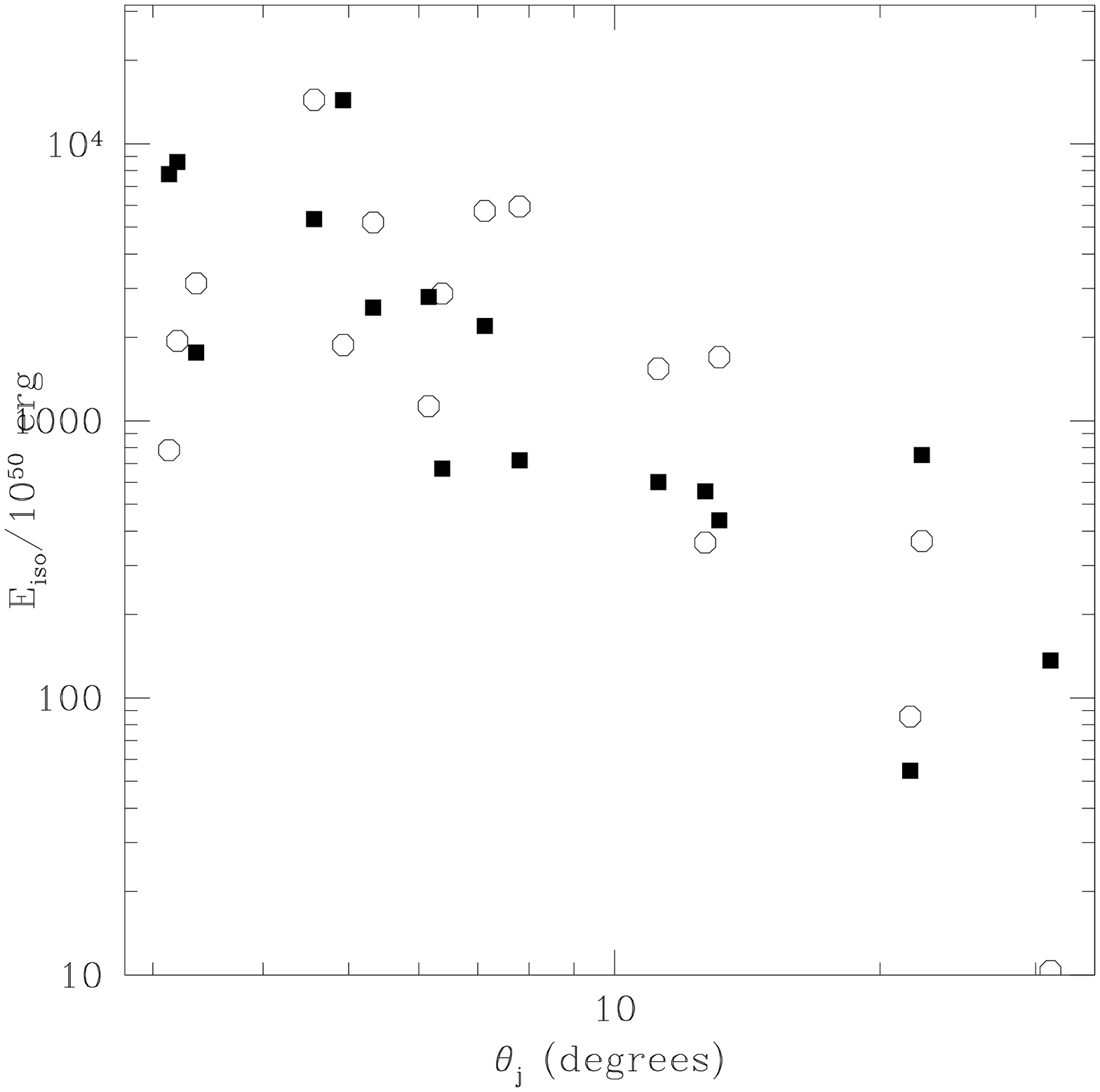} \caption{  Same
as Figure 2, but where the circles are data derived from a Gaussian jet
structure, a normalization $\epsilon_{o}$ that varies as a log-normal with
a mean $<{\rm lg}(\epsilon_{o}/10^{50}{\rm erg})> = 3.4$ and a standard
deviation $\sigma_{{\rm lg}(\epsilon_{o}/10^{50}{\rm erg})} = 0.6$, and a
characteristic jet width that follows the relation
$\epsilon_{o}\theta_{o}^{2}=constant$.} \end{figure}

\begin{figure}
\plotone{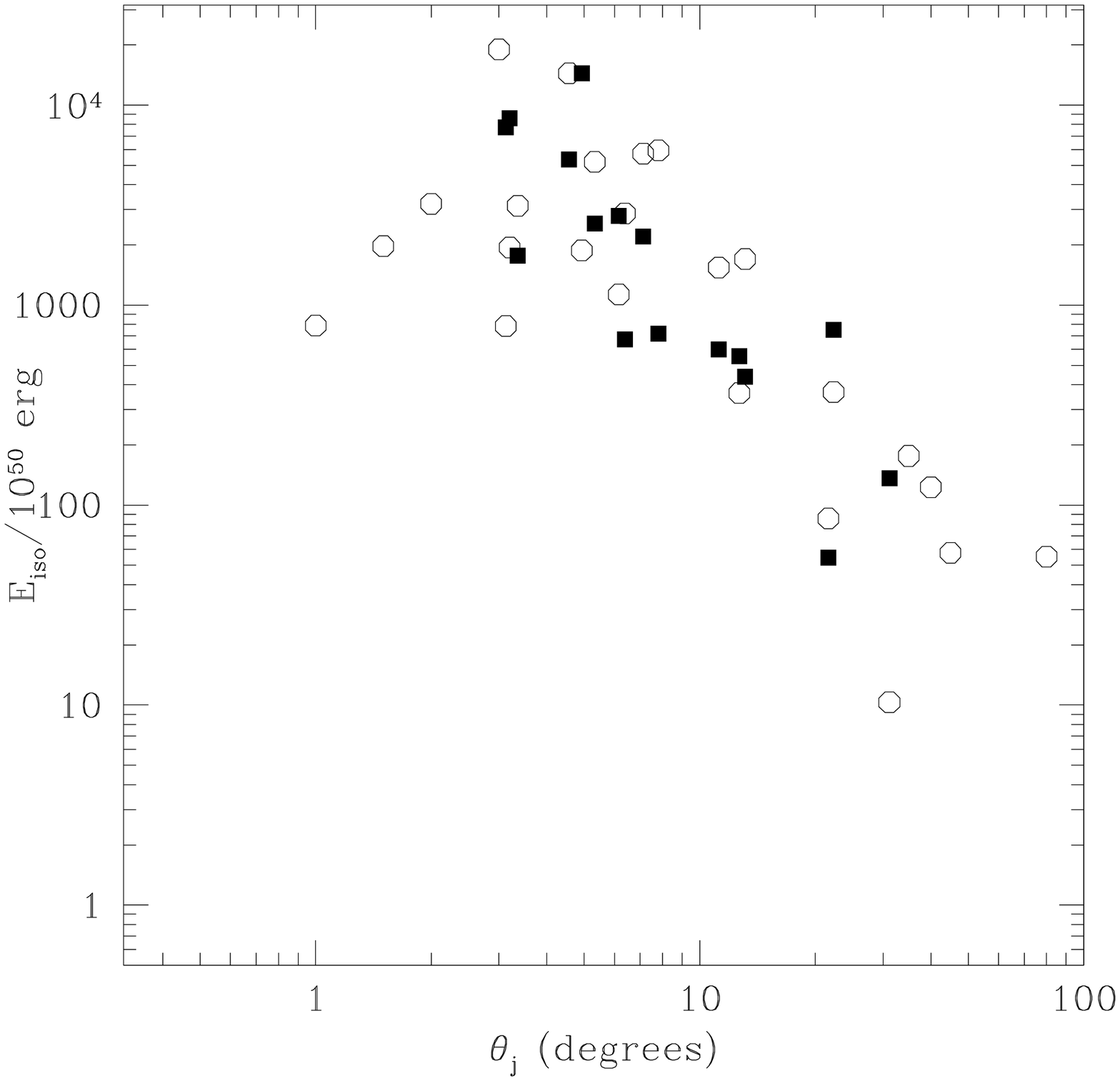}
\caption{Same as Figure 6, but plotted over a larger range in both axes to
show the turnover at low $\theta_{j}$.  For the case of $<\epsilon_{o}>$ 
or $<\theta_{o}>$ varying individually, there is also a sharp cutoff at
high $\theta_{j} \ga 30$ degrees. }

\plotone{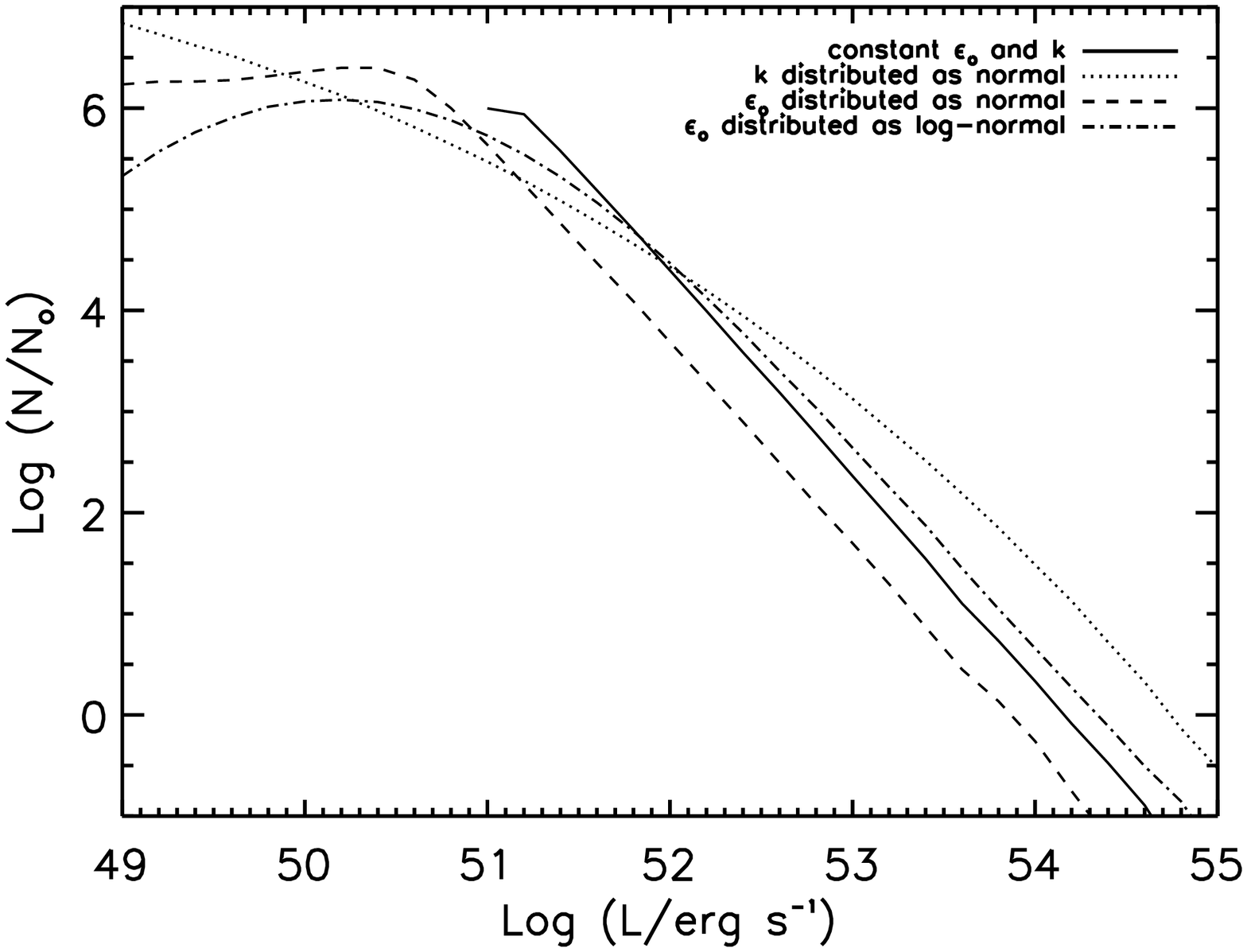}
\caption{Monte-Carlo simulated (differential) luminosity functions for 
GRBs with power-law 
jets, where $N_{o}$ is an arbitrary normalization.
   The luminosity 
function is
calculated assuming isotropic emission and a typical duration to all
bursts.
 The solid, dotted, dashed, and dash-dotted lines represent 
  power-law jets with no variations, with powerlaw index $k$ 
varying as according to a normal distribution, with $\epsilon_{0}$ varying as 
according to a normal
distribution, and with 
$\epsilon_{0}$ varying as a log-normal, respectively. \label{fig:lf1}}
\end{figure}

\begin{figure}
\plotone{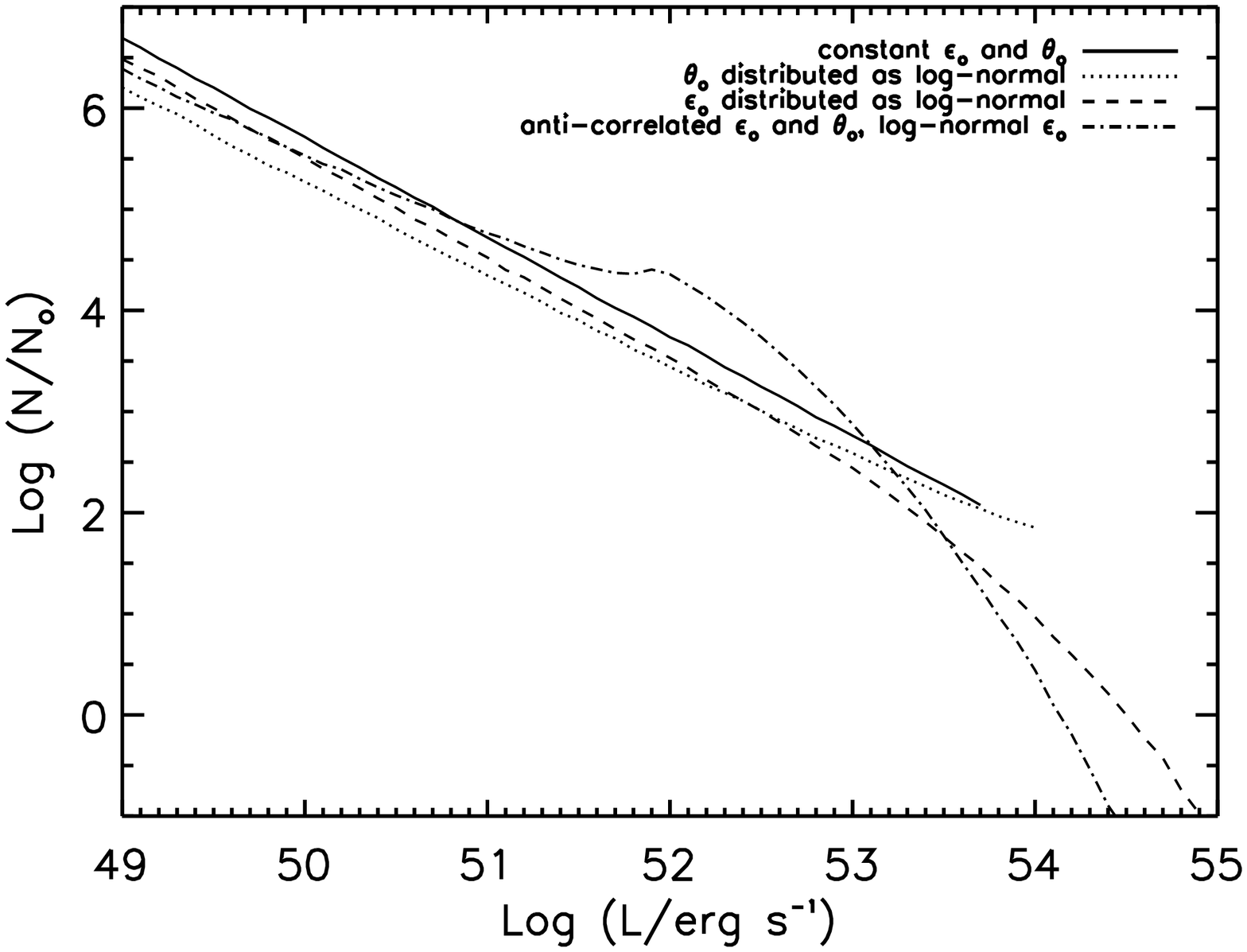}
\caption{Same as Figure 8, but for Gaussian 
jets and log-normal parameter variations.  The solid, dotted, dashed, and 
dash-dotted
lines represent Gaussian jets with no variations, 
with
$\theta_{0}$ varying, with $\epsilon_{0}$ varying, 
 $\epsilon_{0}$ and $\theta_{0}$ varying in an anti-correlated way, respectively.
 \label{fig:lf2}}
 
\plotone{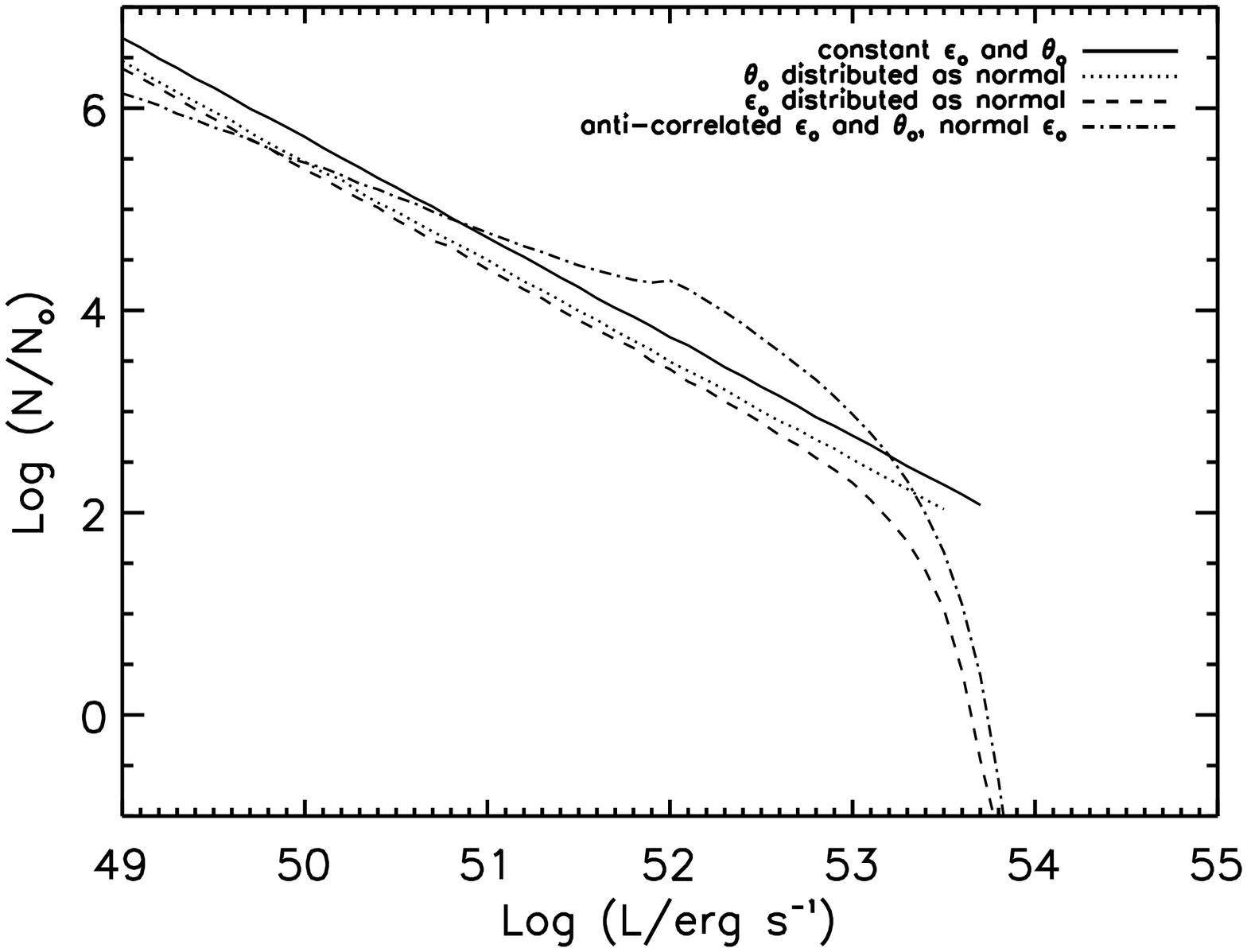}
\caption{Same as Figure~\ref{fig:lf2}, but with the parameters varying according
to a normal distribution. \label{fig:lf3}}
\end{figure}

\end{document}